# The emergence of knowledge exchange: an agent-based model of a software market

Maria Chli and Philippe De Wilde

**Abstract**

We investigate knowledge exchange among commercial organisations, the rationale behind it and its effects on the market. Knowledge exchange is known to be beneficial for industry, but in order to explain it, authors have used high level concepts like network effects, reputation and trust. We attempt to formalise a plausible and elegant explanation of how and why companies adopt information exchange and why it benefits the market as a whole when this happens. This explanation is based on a multi-agent model that simulates a market of software providers. Even though the model does not include any high-level concepts, information exchange naturally emerges during simulations as a successful profitable behaviour. The conclusions reached by this agent-based analysis are twofold: (1) A straightforward set of assumptions is enough to give rise to exchange in a software market. (2) Knowledge exchange is shown to increase the efficiency of the market.

This work has been carried out as part of the project Digital Business Ecosystem, funded by the $6^{th}$ Framework Programme of the European Commission. Contract Number: IST - 2002 - 507953.
This work was carried out when M. Chli and P. De Wilde were at Imperial College London

# The emergence of knowledge exchange: an agent-based model of a software market

## I. INTRODUCTION

The growth of the Internet as a medium of knowledge exchange has stimulated a lot of scientific interest originating from various disciplines. The willingness of individuals, organisations as well as commercial firms to share information via the Internet has been remarkable. In some sectors like scientific research, the communication of newly acquired knowledge and expertise in a field is considered vital for their advancement. On the other hand, in other sectors, the benefits of such exchanges may not be obvious. For instance, it might even be considered damaging for pharmaceutical companies to make public any innovations generated by their Research and Development (R&D) process. In spite of this view, exchange of intellectual property in some industries occurs quite frequently and in various different ways. These include the forming of strategic partnerships, the participation in open source software projects and the publication of scientific papers by research labs that are part of commercial companies.

We study the knowledge exchange that occurs in the software industry. In particular, we focus on analysing the rationale behind this exchange as well as its effect on the industry. The complexity of software requirements is a characteristic that distinguishes the software market from others. However, the findings of this work might be relevant to other industries as well. This effort fits within the framework of the Digital Business Ecosystem (DBE) project. The DBE project is an attempt to develop a distributed environment which will interlink European Small and Medium Enterprises (SMEs) that are software providers and foster collaboration between them.

Our broader interest lies in understanding the dynamics of ecosystems [11], [15], [43]. Furthermore, we are interested in analysing the global system properties which emerge from the interactions that occur in a market ecosystem. We have been using techniques from agent based modelling to simulate the DBE environment. The main aspects of the DBE market are captured in a model where the SMEs are agents with bounded rationality. This model is then studied using simulations of various settings, and a number of observations are made. One of the most interesting observations is that exchanges between the agents similar to the ones that happen in real-life *arise* in the system. This behaviour *emerges* in the market even though the model does not explicitly account for social issues of trust, network effects or managerial strategies.

The paper is organised as follows. The following section gives an insight to the Digital Business Ecosystem project and the characteristic of the market that will be developed. In section III we sketch the background of this work, namely we review the types of exchanges that occur in markets, giving particular attention to the software market. Section IV details the model used for the investigation carried out. Section V analyses the experiments performed and the results produced and section VI concludes.

## II. DIGITAL BUSINESS ECOSYSTEM

In this section we give a brief overview of the Digital Business Ecosystem project, highlighting its aims and motivation. The characteristics of the end-product are identified and special attention is given to the efficiency of the market that will be formed.

### A. A DBE Economy

It is stated in [35] that virtual organisations make dynamic coalitions of small groups possible. In this way the companies involved can provide more services and make more profits. Moreover, such coalitions can disband when they are no longer effective. At present, coalition formation for virtual organisations is limited, with such organisations largely static.

The overall goal of the DBE project [1] [13] is to launch a new technology paradigm for the creation of a digital business ecosystem that will interlink SMEs and especially software providers. The project is encompassed by the European Union's initiative to become a leader in the field of software application development and to strengthen its SME industry. An open source distributed environment will support the spontaneous evolution, adaptation and composition of software components and services, allowing SMEs that are solution and e-business service providers to cooperate in the production of components and applications adapted to local business needs. This will allow small software providers in Europe to leverage new distribution channels providing niche services in local ecosystems and extending their market reach through the DBE framework. Easy access and large availability of applications, adapted to local SMEs, will foster adoption of technology and local economic growth. It will change the way SMEs and EU software providers use and distribute their products and services.

The main objective of this work, which was carried out as part of the DBE project, was to study the properties of this new type of market. It is clear that the interactions and exchanges between the SMEs within the Digital Business Ecosystem environment will have an effect on the dissemination of information and subsequently to the efficiency of the market.

### B. Market Efficiency

Within the environment of the DBE, business alliances, networks and supply chains require much less effort to be formed. This will promote cooperation and easier dissemination of information between the member SMEs. On the other hand, competition for a share of the market between SMEs will become more direct. It is to be hoped, that these factors will raise the levels of efficiency in the DBE market in comparison to a traditional market. While these aspects of the DBE are very interesting and the subject of future research, this work studies how market efficiency is affected by the exchange of information between SMEs. The experiments carried out on our model, confirm that as the agents engage in more information exchanges between them, with time the market efficiency of the system rises.

---
[1] The web page of the project can be found at www.digital-ecosystem.org

Efficient Markets Theory, as proposed by [19], is a field of economics which seeks to explain the operation of an asset market. Specifically, it states that at any given time, the price of an asset reflects all available *information* [3], [12]. The efficient market hypothesis implies that it is not generally possible to make above-average returns in the stock market over the long term by trading lawfully, except through luck or by obtaining and trading on inside information.

The DBE environment is different from an asset market, so the definition of efficiency needs to be modified, retaining the spirit of the efficient market hypothesis. In the model of the DBE used in this work, the market is driven by demand which is fixed and unaffected by the supplied DBE services. In this case the market is efficient if, at any given time, the supply of a service reflects all available information. This means that, the services supplied are such that they satisfy the underlying market needs optimally. In other words, the SMEs are not concentrating on catering for some needs while others are left unsatisfied. In an efficient DBE market, *all the needs will be satisfied evenly, assuming that there is equal demand for each of them*. To draw a parallel between the traditional definition of an efficient asset market and the proposed definition for the efficiency of the DBE market consider the following. In an inefficient asset market, a trading agent can earn excessive returns by buying a particular stock which she believes to be undervalued. Similarly, in an inefficient DBE market a company might make excessive profits by satisfying a need which it knows is not sufficiently satisfied. To invert the argument, in an efficient asset market, asset prices adjust instantaneously and in an unbiased fashion to publicly available new information, so that no excess returns can be earned by trading on that information. Similarly, in an efficient DBE market, the supply of services will adjust immediately to any arising information about the underlying needs.

Cooperation, symbiosis [16], [27] as well as the efficiency [37], [40] of adaptive multi-agent systems has been studied in the context of the simple games. In [40] no verifiable definition of efficiency is given, whereas in [37] the system is considered to be in an efficient market phase when all information that can be used by the agents' strategies is traded away, and no agent can accumulate more points than an agent making random guesses would. In the work presented in this paper, market efficiency, cooperation and competition are studied in the context of a more realistic economic market.

## III. BACKGROUND

In this section we list a number of ways in which exchange of knowledge between companies happens in a market and the rationale for each of them is briefly reviewed. As this work focuses on SMEs that are software providers, we survey the key characteristics of the software industry and the exchanges in this particular market.

### A. Exchange in economic markets

In an economic market there are many ways in which the firms engage in exchanges between them. These include the forming of strategic partnerships, the participation in open source software projects and the publication of scientific papers by research companies like HP Labs and Microsoft Research. In the paragraphs that follow we will briefly examine the rationale behind these different forms of exchange.

For a strategic partnership to be formed, the partners must mutually benefit from the experience, expertise and talent that all the parties bring to the partnership. There usually is an immediate worthy goal or objective that the partners concerned wish to achieve. For instance, they may wish to operate in a new market, or to bring about a change of leadership in the industry they operate in. Hagedoorn in [24] reports a dramatic rise especially in R&D partnerships, over the past 40 years. These partnerships are mostly limited-time project based collaborations as opposed to long-term alliances. The main motives behind them are reported to be related to cost-cutting as well as risk minimisation whilst the partners attempt to enter new technological areas.

Recent economics and management research has studied the phenomenon of commercial firms contributing to open source projects. The main motive indicated by these analyses is strategic [22], as set out in more detail in section III-B where the specifics of the software industry are analysed. This seems to be consistent with the fact that it is not the leaders in the industry who engage in open source development, but the followers.

Another form of exchange, which at first might seem counter-intuitive, is the publication of scientific papers containing the findings of the research commercial companies perform. It may be argued that it would be in the interest of those companies, to keep their innovative work to themselves. Another argument, however, is that by publicising their research they invite others to endorse it, add to it and in effect advance it further. Then, they can use the knowledge acquired by this process to better their products.

The model of a software market that we propose as part of this work is simple in the sense that the agents/firms do not have the ability to reason about complex situations. They cannot make decisions to operate in new markets, or form partnerships in order to change the leadership in the industry. They cannot devise strategies to undercut their competitors. However, they operate in a capitalistic economy where the best of them succeed whilst the worst perish. They are thus equipped with a simplistic mechanism of reinforcement learning, i.e. being rewarded or punished for choices that prove to be good or bad respectively. When given the opportunity to engage in exchange of services between them, they learn with time under which circumstances this is beneficial to them and they proceed with it without ever being biased by external factors towards exchanging.

### B. The software industry

Complexity is a key characteristic of software which distinguishes the software industry from others. Typical software products carry a large number of features, with innumerable [2] interactions between them. For a program to be successful in the market, it is necessary that it has the right set of features to satisfy the customer base and that these features operate successfully together.

The market of proprietary software providers/publishers is dominated by large companies, not SMEs. Microsoft Corporation holds the lion's share in the software market with companies like Oracle, IBM, Hewlett-Packard and Sun following with smaller shares[2].

At the same time, the open source[3] movement has been quite successful in developing relatively complex software products like Linux, Apache or sendmail that are serious competitors of well established proprietary software [38]. Networks of thousands of volunteers have contributed to these highly complex products. This appears, as it is pointed out in [2], to counter the economic intuition that private agents, without property rights, will not invest sufficient effort in the development of public goods because of free-rider externalities.

Lerner and Tirole in [33] justify the volunteers' motivation for contribution to the open source movement as an opportunity to 'signal their quality'. In other words, the volunteers believe it will enhance their career prospects, as the names of the contributors are always listed in open source projects. Other individual motivations, like altruism or opportunity to express creativity are also mentioned.

It is important to point out that in recent years, open source projects have not only received contributions by individuals. There have been organised efforts by firms like Sun, IBM and others that have endorsed such projects. The survey [6] conducted among firms, as well as the account of [20] of Sun Microsystems and [22] list strategic reasons behind the motivation of firms to contribute to open source projects. These reasons include efforts to undercut rival products, gaining a wider tester base for their own products, initiating a gift economy culture between the firm and the open source developer community (where the firm provides the software for free and the community provides debugging or more source code in return) and giving out the software to clients in order to charge for its maintenance and support.

---

[2]The information reflects the year 2002-2003 and was obtained from IBIS World, a strategic business information provider. http://www.ibisworld.com/snapshot/industry/default.asp?page=industry&industry_id=1239 accessed on 27/05/2005.

[3]In open source software, the source code for a program is made open and available for anyone to screen. There are different open source licenses which prescribe what one is allowed to do with the source code e.g. screen it, interpret it, make changes etc. This is in contrast to proprietary software licenses where the source code is protected by property rights against modification.

Previous work in this area includes that of Johnson in [28] and Bessen in [2] who have used mathematical models to explain the emergence of the open source initiative. Johnson focuses more on analysing the individual motives and establishing the relationship between the size of the developer base and whether the development goes on. On the other hand, Bessen concentrates on the firm motives for participation in open source initiatives. Bessen, models software as a bit string, each bit being a certain feature of the software. In this way the notion that the number of combinations of features grows exponentially with the number of features is captured, depicting the complexity the software can have. In his work, he compares open source development with proprietary, pre-packaged provision of software and concludes that the two complement each other, recognising that they serve different groups of customers. The latter suits customers with standard, non-complex software needs, while the former serves customers who have software development capabilities and who need more complex software products.

Bonaccorsi and Rossi in [5] have designed a multi-agent system simulation with which they explore the circumstances for adoption of open source software. They also conclude that proprietary and open source software will coexist in the future. Their model of the diffusion of the two competing streams of software production takes into account issues like the effect of advertising, network externalities and achievement of critical mass as in [34].

The stylised model presented in this work simulates a market in which the companies try to satisfy a set of underlying software needs with the services that they develop. The companies follow simple, high-level rules imposed by a capitalistic economy. Interestingly, exchanges between the agents similar to the ones that happen in real software markets, arise in the system. This behaviour *emerges* in the system even though we have avoided modelling issues like social or strategic motives of the contributors or network effects.

## IV. AN AGENT-BASED MODEL OF THE DBE

### A. Agent-based Modelling

Agent-based modelling has been recently used in Economics research work to study models of markets, e.g. the Santa Fe artificial stock market [4], [32], and their characteristics [31], in Computing-Economics interdisciplinary work to study information economies of autonomous agents [14], [23], [29], [30], [39] and business processes [26], in Social Sciences to study emergent behaviour [17], issues of trust [18] and to perform syndromic behaviour surveillance [10] and in other disciplines.

Much research in multi-agent systems explores how refinements to one agent's reasoning can affect the performance of the system [8]. Significant effort has been directed towards formally defining emergence in agent-based systems. A strong emergent property is a property of the system that cannot be found in the properties of the system's parts or in the interactions between the parts [1]. Additionally, in [42] the notion of universality is studied: systems whose elements differ widely may have common emergent features.

Agent-based modelling according to [41] "is a method for studying systems exhibiting the following two properties:

1) the system is composed of interacting agents; and
2) the system exhibits *emergent* properties, that is, properties arising from the interactions of the agents that cannot be deduced simply by aggregating the properties of the agents."

In models like the one proposed below, where the interaction of the agents is determined by past experience and the agents continually adapt to that experience, mathematical analysis is typically very limited in its ability to derive the dynamic consequences. In this case, agent-based modelling might be the only practical method of analysis.

We follow a 'bottom-up' approach, after a brief overview of the methods used in section VII which follows, in sections IV-B and IV-C we describe the first principles of agent behaviour and in section V we analyse the macro-properties emerging from the agent interactions.

### B. The setting

In this section, the model used for the simulation of the DBE environment is set out.

SMEs are modelled as agents in a multi-agent system. The services the SMEs provide are modelled as bit strings in the same manner software services are modelled in [2], each bit symbolising a feature of the service. Finally, the underlying market is modelled by a set of requests (market needs) which are exogenous and are generated randomly. A request is a bit string of the same size as a service bit string.

Each SME has a population (or portfolio) of services. This population is not static throughout the lifetime of the SME. If a service is successful, the SME tends to add similar services to the portfolio while an unsuccessful service is usually discarded. The whole process is modelled quite elegantly by a genetic algorithm (GA) within the portfolio which involves mutation and crossover with survival of the fittest. Through this population each SME can choose which request it will try to satisfy. The genetic algorithm represents the R&D businesses perform in order to improve their services. An overview of genetic algorithms is given in appendix VII.

The use of genetic algorithms is a natural and simple way to model R&D, with minimal assumptions. The GA captures the following characteristics:

1) trying to find a solution to a particular problem,
2) using a population of possible solutions.

Any other method that can capture the above two characteristics may be used in place of the GA.

The objective of an SME is to increase its fitness. Each SME maintains a portfolio of candidate services, only one of which will be submitted to the market. Each candidate service receives a rating according to how profitable it would be for the SME if it was submitted to the market. This calculation is performed using the services submitted by all other SMEs in the previous round. The rating of each candidate service within the SME portfolio is used to: a) decide on which service to submit to the market and b) evolve the best services in the portfolio (with mutation and crossover) and eliminate the worst services.

The fitness of a service measures how profitable it is to its owner. The profitability of a service depends on:

1) how close the service is to the market needs (service-request similarity) and
2) how many other services satisfy those needs (limited demand).

The fitness of an SME equals the fitness of the service it offers.

In the section that follows we discuss the factors that affect the fitness (or profitability) of a service.

*1) Service-Request Similarity and Limited Demand:* Assume there are $m$ SMEs in the market, each one offering a single service. Consider a service S and a request R, each represented by a bit string of fixed length. Similarity is measured by the percentage of shared bit values between S and R, denoted by $d(R_i, S_j), 0 \leq d \leq 1$. If the market requests are $R_1, R_2, ..., R_n$, services in the market are $S_1(t), \ldots, S_m(t)$, the fitness of a service $S_j(t)$ is

$$U_j(t) = \sum_{i=0}^{n}(\phi(R_i, S_j(t)) \times \rho_i(t)), \quad (1)$$

where

$$\phi(R_i, S_j(t)) = e^{-\frac{1-d(R_i, S_j(t))}{\alpha^2}}. \quad (2)$$

The variable $\phi$ is used to parametrise the fitness landscape (make maxima more or less pronounced), $\alpha$ being a shape parameter. Figure 1 shows the relationship of $\phi$ with with the similarity $d$. The weight/discounting factor $\rho$ is given by

$$\rho_i(t) = min\left\{1, \frac{1}{\sum_{j=1}\phi(R_i, S_j(t))}\right\}. \quad (3)$$

The variable $\rho$ models the fact that the demand in the market is limited. When a request is saturated (i.e. too many services try to satisfy it) then $\rho < 1$. Subsequently, the fitness of the service is discounted. Otherwise, when $\rho = 1$ the fitness of the service equals $\phi$.

The fitness of an SME is equal to the fitness of the service it submits to the market.

*2) Satisfaction of Requests and Market Efficiency:* An additional useful measure is the degree to which a request is satisfied. This is a metric of how saturated it is, in terms of how many services try to satisfy it and how similar their features are to those of the request. The degree of satisfaction $Q_i(t)$ of a request $R_i$ at round $t$ is given by:

$$Q_i(t) = \sum_{j=1}^{m} \phi(R_i, S_j(t)). \quad (4)$$

This measure is necessary for assessing the efficiency of the DBE market. As discussed in section II-B, in an efficient DBE market all the market requests will be equally saturated, assuming there is the same demand for all of them. Thus, we calculate the standard deviation $\sigma(t)$ of the satisfaction values of all the requests in the market at round $t$. The smaller it is, the more similar to each other the saturation levels of the requests are.

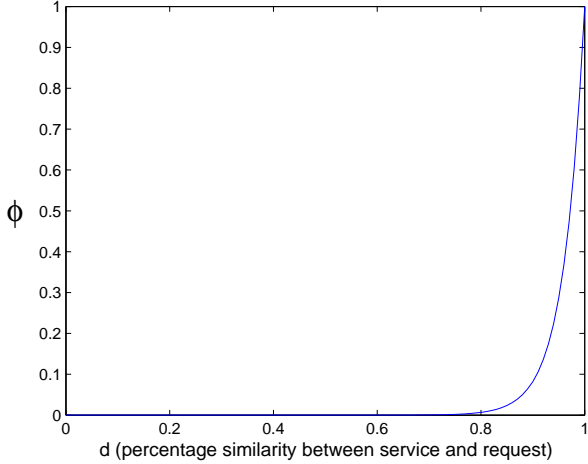

Fig. 1. The relationship of $\phi$ with the service-request similarity $d$ for $a = 0.2$. The variable $\phi$ is used to parametrise the fitness landscape (make maxima more or less pronounced).

$$\sigma(t) = stdev\{Q_1(t), \ldots, Q_n(t)\} \quad (5)$$

The mean of the saturation values will be constant due to the demand in the model being fixed.

### C. Exchange of Services

As outlined in III-A exchange of services may encompass many real-life situations that occur in a market. These include the forming of strategic partnerships of companies, participation in free/open source projects and others. The setting described here is a loose model of such situations which aims to identify the basic factors that lead to this general behaviour of exchanging.

In our model, the exchange involves selecting a set of services from one SME's portfolio and swapping them with the corresponding set of services of the other SME's portfolio. When a company chooses to swap a set of services, this means that after the exchange has taken place it won't have these services in its portfolio any more. The services in a portfolio of a company are sorted according to their fitness (i.e. how profitable they are to the SME that owns them). The model in its current state supports exchange of services that are in the same rank, in the two portfolios, e.g. the $5^{th}$ service in the portfolio of one SME with the $5^{th}$ service in the portfolio of the other[4].

At each time tick, the SMEs need to decide whether they want to exchange some of their services with one of the other SMEs. A statistical classification algorithm is used to model the decision problems an individual agent faces. An overview of statistical classification is given in Appendix VII.

[4]Experiments have shown that the rank of the services being exchanged is not of much significance, assuming that services of the same rank are being exchanged, but we plan to investigate this further in the future.

*1) Exchange decisions:* Every SME has a classifier system which it uses to decide on whether they want to exchange some of their services with one of the other SMEs. The rules of the classifier are shown in table I below. The objective of an SME at all times is to increase its fitness.

The rules' condition part refers to the rank of the SME in the market with respect to the rank of its colleagues. The action part examines the potential partner's rank and prompts the SMEs either to engage in an exchange with a specific type of partner or abstain from exchanging. For simplicity, the SMEs are clustered in three[5] groups according to their rank. Therefore we have upper, middle and lower ranked SMEs. For an exchange to take place both parties need to agree.

We experiment both with settings in which the rank is based on the fitness of each company and others where the rank is not linked to SME performance in any way. For example, in experiments where rank is based on SME performance, the SME with the highest fitness will have $rank = 1$, whilst the SME with the lowest fitness will have $rank = number\ of\ SMEs$. On the other hand, in experiments where rank is unrelated to performance in the market the rank of an SME may be its id number. In section V we analyse these experiments and present the effect the different meanings rank may take have on the learning that occurs.

| if | my rank = lower | then | exchange with lower cluster, | $s_1$ |
|---|---|---|---|---|
| if | my rank = lower | then | exchange with middle cluster, | $s_2$ |
| if | my rank = lower | then | exchange with upper cluster, | $s_3$ |
| if | my rank = lower | then | do not exchange, | $s_4$ |
| if | my rank = middle | then | exchange with lower cluster, | $s_5$ |
| if | my rank = middle | then | exchange with middle cluster, | $s_6$ |
| if | my rank = middle | then | exchange with upper cluster, | $s_7$ |
| if | my rank = middle | then | do not exchange, | $s_8$ |
| if | my rank = upper | then | exchange with lower cluster, | $s_9$ |
| ⋮ | ⋮ | | ⋮ | ⋮ |

TABLE I
A FEW EXAMPLE RULES OF THE CLASSIFIER WHICH AN SME USES TO DECIDE ON WHAT TYPE OF PARTNER TO CHOOSE FOR AN EXCHANGE.

The classifier system operates as follows [31]. First, it examines the `if` part of each rule to determine and shortlist the rules whose conditions are satisfied at a given time t. It then assigns a score $b$ to the shortlisted rules, $s_k$ being the strength of the $k^{th}$ rule:

$$b_k(t) = s_k(t) + \varepsilon, \text{where } \varepsilon \simeq N(0, \sigma). \quad (6)$$

The rule with the highest score $b$ becomes the *active rule*.

After the active rule has been executed and has generated payoff $\omega$ during the previous round $t-1$, the classifier system updates its strength $s$:

$$s_k = s_k(t-1) - cs_k(t-1) + c\omega(t-1), \text{where } c \in [0, 1]. \quad (7)$$

In other words, $\Delta s_k(t) = c[\omega(t-1) - s_k(t-1)]$. Therefore, as long as the payoff in round $t-1$ is greater than the strength of the rule on that round, the strength will increase. If the selection of the rule led to a small payoff being generated, the strength of the rule will decrease, making it less likely to be activated in the future. The strength of each rule converges to some weighted average of the rewards $\omega$ generated by the environment in response to that specific rule.

[5]Experiments have been carried out which showed that model behaviour doesn't vary significantly with cluster size. Three is the optimal number of clusters with respect to having a model which is realistic enough while taking a reasonable amount of time to execute and giving us the ability to present the results in an efficient and clear way.

In our implementation of the model all the rules have initial strength 0. The rule strengths are adjusted as the simulation goes on. The strength of each rule that is activated is updated at every round using the following payoff from the external environment: $\omega(t) = U_j(t) - U_j(t-1)$. In other words, the payoff is the difference in the fitness of the company between the current and the previous round. The payoff may be negative, zero, or positive according to the change in fitness.

*2) Exchange decisions resolution:* Once the companies that have decided to participate in an exchange have selected the type of partner they prefer, they are teamed up accordingly. For instance, an SME in the cluster of middle ranked SMEs, who has decided to exchange with a high fitness company will be coupled with a high ranked company who wants to exchange with a middle ranked one. If a suitable partner is not found the exchange does not happen. The strength of the rule that was activated in that case will still be updated even if the transaction was not carried out. This reflects the effect choosing a partner who is unwilling to collaborate has on the fitness of the company.

*D. Discussion*

The model outlined above is simple in that it has captured the main aspects of a digital business ecosystem. It is the model of a market in which the companies try to satisfy a set of underlying requests. They do so by producing and making available services that are as close as possible to the specified requests. Each company has its own R&D portfolio of services that it evolves. At each round the companies go to the market with what they believe is the best service in their portfolio. In addition, the companies have an option to exchange services with partners that they select themselves.

The simplicity of the model is also inherent in the behaviour of the agents. The agents have to find which is the best service to make available, based on the services that were submitted to the market during the previous round. Also, they need to decide whether and with whom to exchange their services based on their rank in the market. These are all abstractions from reality. We do not assume any network effects in the market. Also, there are no indicators about value of the brand of a company.

V. ANALYSIS OF THE MODEL

In this section the experiments carried out using the model of the DBE are described. The analysis focuses on two main findings:

1) The companies discover themselves that under certain circumstances it is beneficial to them to exchange services between them.
2) Allowing exchange to take place in the market, makes for greater market efficiency levels.

It is important at this point to stress that the choice to exchange services is not a practice that is imposed by the model mechanism. Instead, it is a feature that emerges from the classifiers as it is a gainful practice for the companies under certain circumstances.

The model behaviour is quite general and has been observed for a very wide range of parameters and initial conditions. The graphs and figures shown below come from randomly selected runs of the simulation, unless it is stated otherwise.

*A. Service Exchange*

*1) Exchange Decision:* As described in section IV-C each agent/company uses a classifier to decide whether or not to exchange some of its services. The decision is based on the company's rank in the market. Figures 2(a) and 2(b) show the average strength of the rules of all the companies' classifiers at the end of a simulation which lasted for 10 000 iterations. The companies are ranked according to their fitness. The fittest company will have rank 1 whilst the least fit company will have rank equal to the number of companies in the market. To make for less time consuming simulations and more readable graphs the companies are grouped into three clusters according to their rank; so they are divided into lower, mid and upper ranked SMEs. Figure 2(a) was generated from a run of the simulation where the DBE market consisted of 21 SMEs, each having 20 services in its portfolio. Each service had 10 features. There were 4 software requests in the market, generated randomly. The run of the simulation which produced figure 2(b) had largely similar parameters, the difference being that there were 30 services in the SMEs' portfolios and there were 5 requests in the market.

The strongest of the rules at each situation is the one which is more likely to be activated. In other words, it is shown in figures 2(a) and 2(b) that if a company belongs to the mid or lower cluster it is likely that it will choose to participate in an exchange (preferably with a upper ranked company) while if it belongs to the upper ranked cluster it will avoid engaging in any exchange activities. The graphs show that in the less successful, lower ranked SMEs the classifier rules that correspond to `exchange` actions have higher strengths than the rule that leads SMEs not to exchange. The opposite holds for higher ranked SMEs, i.e. the rule that corresponds to a `not exchange` action has higher strength than the `exchange` rules. For mid-ranked SMEs, a rule prompting the firm to exchange is the stronger of all, but exchanging is not always a profitable practice; the rule that leads the SME to avoid exchanging is often stronger than some `exchange` rules.

The generality in the behaviour of the model is confirmed by figure 2(c). A wide range of parameters and initial conditions were varied in a total of 200 experiments, keeping the number of SMEs in the market constant (21). Figure 2(c) shows the average values of the SME classifiers' strengths over those 200 experiments. The general trend which emerges is that the average performing (mid cluster) and worst performing (lower cluster) SMEs learn that it is to their advantage to exchange services with others while the top performers (upper cluster) learn to avoid exchanging .

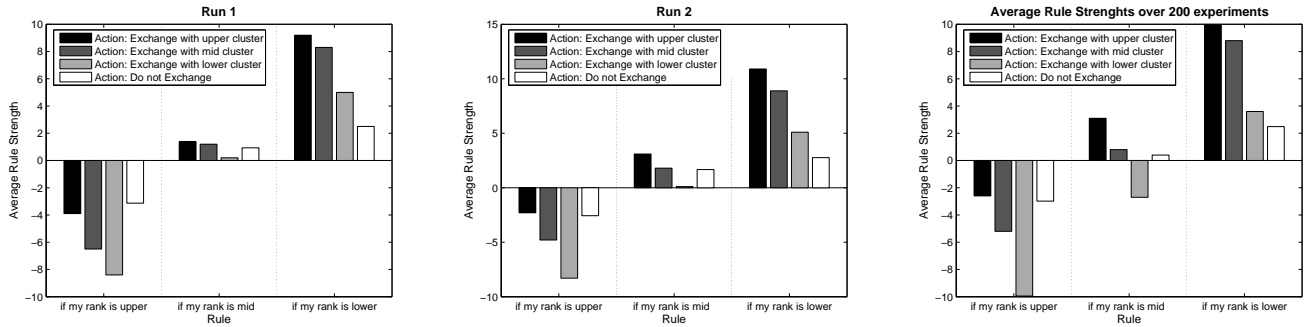

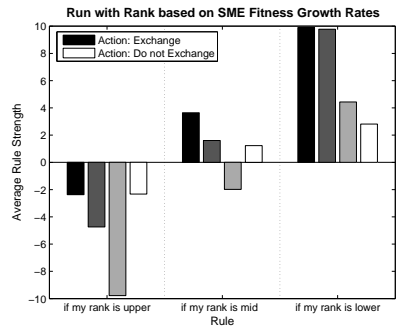

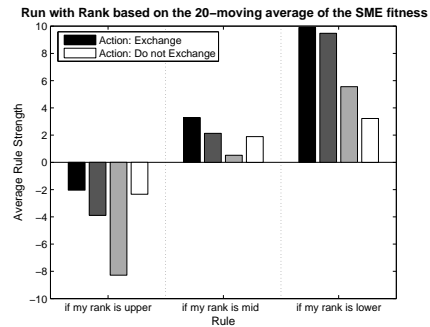

(a) Run 1     (b) Run 2     (c) Average over 200 Experiments

(d) Rank based on Fitness Growth Rate     (e) Rank based on Fitness Moving Average

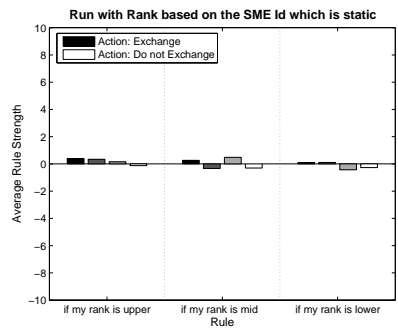

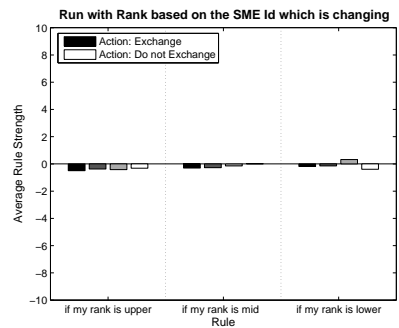

(f) Rank based on SME Id which is random and static throughout the simulation

(g) Rank based on SME Id which is random and constantly changing throughout the simulation

Fig. 2. **Average Exchange Rule Strength** The graphs show the strength values of each rule at the end of a simulation averaged out over all SMEs' classifiers. The SMEs decide whether to participate in an exchange of services according to their rank. The classifier each SME has is as follows:

| if | $my\ rank = lower$ | then | exchange with lower cluster, | $s_1$ |
| if | $my\ rank = lower$ | then | exchange with middle cluster, | $s_2$ |
| if | $my\ rank = lower$ | then | exchange with upper cluster, | $s_3$ |
| if | $my\ rank = lower$ | then | do not exchange, | $s_4$ |
| if | $my\ rank = middle$ | then | exchange with lower cluster, | $s_5$ |
| ⋮ | ⋮ | | ⋮ | |

For figures 2(a)-2(e) the rank of the SMEs is based on measures related to their fitness, while figures 2(f) and 2(g) were created for settings in which the SME rank was unrelated to fitness. The graphs show in settings where the rank is associated with some fitness measure the SMEs that are further down in the rank learn that is beneficial to them to participate in an exchange.

**2(a) Run 1** parameters: 21 SMEs, each having 20 services in its portfolio. Each service had 10 features. There were 4 software requests in the market. The rank was based on the fitness value of the SME.

**2(b) Run 2** parameters: 21 SMEs, each having 30 services in its portfolio. Each service had 10 features. There were 5 software requests in the market. The rank was based on the fitness value of the SME.

**2(c) Average values over 200 experiments** This figure confirms the generality of the behaviour of the model. A wide range of parameters and initial conditions were varied in a total of 200 experiments, keeping the number of SMEs in the market constant (21). The rank was based on the fitness value of the SME.

**2(d) and 2(e) Average Exchange Rule Strength based on SME performance measures.** The SMEs decide whether to participate in an exchange of services according to their performance. In 2(d) the performance measure deciding the rank of the SMEs is their fitness growth rate, while in 2(e) it is the 20-moving average of the SME fitness. When the ranking of the SMEs is performance related information exchange emerges as a gainful strategy.

**2(f) and 2(g) Average Exchange Rule Strength not based on SME performance measures.** In 2(f) the SMEs decide whether to participate in an exchange of services according to their unique id. In 2(g) the ranking of the SMEs is random and constantly changes. In both cases, the ranking is unrelated to SME

To understand better the behaviour of the system we performed experiments with different rankings of the SMEs. Amongst the ranking methods we tested were variants of the fitness ranking, as well as rankings unrelated to SME performance altogether. The results seem to indicate that information exchange emerges as long as the ranking is in some way related to SME performance. We show in figure 2(d) the rule strengths in the case the SMEs were ranked according to fitness growth rates

$$\Delta U_j(t) = U_j(t) - U_j(t-1), \quad (8)$$

rather than fitness itself. The graphs produced are similar in pattern to those in figure 2(c). These strengths imply that the rules are significant and learning has taken place in the system. Similar results, shown in figure 2(e), were produced when SMEs were ranked according to the N-moving average of their fitness, given by

$$\mu = \frac{1}{N} \sum_{T=t-N}^{t} U_j(T). \quad (9)$$

On the other hand, in figure 2(f) a typical case of a ranking that is unrelated to SME fitness is shown. In that particular case we gave the SMEs an arbitrary ranking that remained fixed throughout the simulation. The rule strengths indicate that no rule is significantly more important than any other one implying that the rules are not relevant and no learning has occurred. We also tried a completely random and constantly changing SME ranking which produced similar results, shown in figure 2(g).

*2) Choice of Exchange Partner:* An interesting result which arose from the experiments is the choice of potential partners for the companies who decide to exchange. In all three situations (`if my rank is upper`, `if my rank is mid` and `if my rank is lower`) the strength of the rules that prompt SMEs to exchange reveal a decreasing preference from left to right between upper, mid and lower ranked partners. That result is entirely intuitive and confirms the validity of the model.

A result that might not be so obvious is the fact that the lower ranked SMEs benefit from exchanging even between themselves. This is reflected in the fairly high strength of the relevant rule and it is better illustrated in figure 3.

The experiment that yielded figure 3 is as follows. To make for a more intelligible graph, there are only six SMEs in the market and two distinct requests. Every 400 rounds the underlying requests in the market change. Every 200 rounds (but not when the requests change), the lower ranked SMEs exchanged services between them. As the purpose of this experiment was to verify the finding that exchange among lower ranked SMEs is beneficial, the exchange was done deliberately and not using the classifier. As shown in figure 3, in round 200 the exchange does not upset the equilibrium too much as the SMEs have more or less the same fitness. In round 600 the exchange drives the lower ranked SMEs up, whilst damaging the fitness of the others in the market. In round 1000 the exchange not only drives the under-performers up but also causes one of them, $SME_1$ to join the upper cluster.

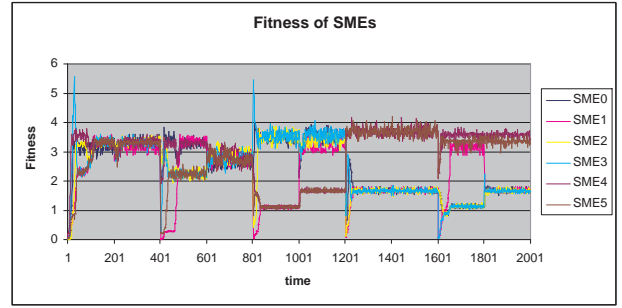

Fig. 3. This is an experiment that illustrates that exchange among lower ranked SMEs is beneficial to them. Every 400 rounds the underlying requests in the market change. Every 200 rounds (but not when the requests change), the lower ranked SMEs exchanged services between them. In most instances the exchange drives the under-performers up, in terms of fitness.

The experiment described above illustrated that exchanges between low-ranked SMEs can be highly beneficial. This is because the fusion of their portfolios might yield services that enable them to operate in a new market segment, in other words it may lead them to satisfy another request which was previously not catered for. This can cause their rank in the market to improve and even bring about a change of leadership in the industry.

### B. Market Efficiency

As discussed in section II-B, the increased flow of information within the DBE, will make it easier for the participating companies to find the right trading partners. Consequently, it will make for greater market efficiency levels in comparison to a conventional market (e.g. the software industry). An interesting observation which emerged from the analysis of the simulations carried out is that allowing the SMEs to exchange services between them, increases the efficiency further.

A DBE market is considered efficient when all the requests are equally saturated. In an efficient DBE market, the supply of services will adjust immediately to any arising information about the underlying requests. In other words, there is no excess profit to be gained by an SME choosing to satisfy another request than the ones it currently does. As mentioned in section IV-B.2, the degree of satisfaction of a request R is given by equation 4. In order to assess the level of efficiency in the market we need to calculate the standard deviation $\sigma(t)$ of the satisfaction values of all the requests in the market, as given by equation 5. The smaller it is, the more similar to each other the saturation levels of the requests are. It is important to mention at this point that the mean of the saturation levels remains constant, because in the model we assume equal demand for all of them, and it is equal to $\frac{\text{number of services in the DBE}}{\text{number of requests}}$.

Figure 4 shows the standard deviation $\sigma(t)$ of the saturation values $Q_i(t)$ of all the requests $\{R_1, \ldots, R_4\}$ in the market, for two different runs of the DBE simulation. Both runs had been initialised with the same parameters, for one of them exchange between the SMEs was not permitted, whereas for the other one the SMEs were free to exchange services with each other according to the procedure detailed in section

IV-C. In order to train the classifiers used for the exchange decisions, every 500 rounds all SMEs' portfolios were reset to the services they had at round 0. To make comparison easier, the resetting of the portfolios was also done during the run where exchange was not allowed. In effect, in this experiment, 'history' repeats itself every 500 rounds. This is the reason spikes occur in the graph every 500 rounds. When exchange is permitted, the SMEs are given the chance to exchange services with each other at rounds 250, 750, 1250, 1750, etc. The graph shows a period of 5000 rounds, when the classifiers have been sufficiently trained.

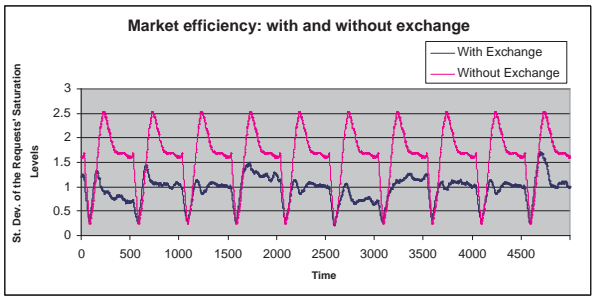

Fig. 4. Market Efficiency: We assess the level of market efficiency by plotting the standard deviation of the saturation degrees of the requests in the DBE Market. The smaller the standard deviation, the greater the market efficiency. The graph contrasts these data for a situation in which the SMEs are allowed to exchange services with each other and for a situation where exchange is not allowed. The standard deviation of the saturation degrees of the requests is significantly smaller when exchange is allowed, indicating a more efficient market. For classifier training purposes every 500 rounds all SMEs' portfolios were reset to the services they had at round 0. In the case where service exchanges are allowed, these happen in the middle of each cycle, i.e. at rounds 250, 750, 1250, 1750, etc.

It is evident from the graph, that when exchange of services between SMEs is allowed, the standard deviation of the requests saturation values is considerably smaller. In other words, the requests in the market are more evenly satisfied. This result is quite invariant to initial conditions and parameters of the simulation. So in the system described, not only will SMEs adopt information exchange as beneficial to their individual progress, but it will also result in a global improvement to the efficiency of the market. Again this is in agreement with what is observed in real economies where open standards, publication of innovations and dissemination of ideas lead to highly efficient markets.

## VI. CONCLUDING REMARKS

The aim of this work has been to study the rationale as well as the effect of knowledge exchange in economic markets. We focus especially on the software industry, our findings, however, to some extent apply to other industries as well. Sharing of information between commercial firms is considered controversial. Although it is acknowledged that when two companies join forces to develop an innovative product they can both benefit, sharing trade secrets is not undertaken lightly. Our main aim has been to formalise a plausible and elegant explanation of how and why companies adopt information exchange and why it benefits the market as a whole when this happens.

An agent based model of a Digital Business Ecosystem market has been implemented to assist us in understanding the dynamics of the market mechanisms. Firms are modelled as agents with minimal reasoning capabilities. We investigated the properties that emerge from the agent interactions that occur in the market. Specifically, we examined two key characteristics that we observed in the simulations carried out. Namely, the fact that the agents discover themselves that under certain circumstances it is beneficial for them to exchange services and that allowing exchange to take place in the market, makes for greater market efficiency levels.

The technologic infrastructure of the DBE will facilitate the dissemination of knowledge among the member SMEs, increasing the volume and the speed of the information flowing in the market. As a result, it is expected that it will allow for greater market efficiency levels in comparison to a conventional market. Admittedly, it is difficult to compare the market efficiency of two different markets. However, an interesting result arose when we performed simulations of the DBE contrasting settings in which exchanges among SMEs were permitted with settings where exchanges were not permitted. Exchanges among SMEs within the DBE further increase the efficiency of the market, which is in agreement with the common intuition that exchanging information is ultimately beneficial for the entire market.

The second and most important conclusion that emerged from the DBE simulation is that exchanges between the agents similar to the ones that happen in real-life arise naturally in our system. At regular time intervals, the SMEs were given the chance to decide whether they wanted to choose a partner and swap some of their services. The decision was taken using classifiers, which were separate for each agent. The agents were not pre-programmed or biased in any way to engage in exchanges. The SMEs, on their own, discovered in which cases exchanging is beneficial for them and what type of partner is the best. Exchange is a practice that emerges, and is not forced upon the agents.

This work does not directly advocate knowledge exchange as a means of increasing profitability of software companies. Knowledge exchange, is indeed an already existing phenomenon in industry as explained in section III-A. The results presented merely serve as a demonstration of a parsimonious set of assumptions that give rise to exchange in a software market. In other words, we identify the substance of this phenomenon, ridding it from unnecessary assumptions, like network effects, social issues of trust, or managerial strategies and show the minimal set of assumptions that allow it to emerge.

## VII. METHODOLOGY: EVOLUTIONARY ALGORITHMS

In order to model evolution in populations as well as learning we have used several evolutionary algorithms in our model. In this section we give a brief overview of these algorithms.

Evolutionary algorithms [7] 'is an umbrella term employed to describe computer-based problem solving systems which use computational models of some of the known mechanisms of evolution as key elements in their design and implementation.' A variety of evolutionary algorithms have been proposed by several researchers. The major ones are: genetic algorithms, evolutionary programming, evolution strategies, classifier systems and genetic programming. They all share a common concept of simulating the *evolution* of objects/structures using the processes of selection, mutation and reproduction. The processes depend on the performance/fitness of the individuals under consideration as defined by their environment and quantified by a fitness function.

More precisely, evolutionary algorithms maintain a population of structures, that evolve according to rules of selection and other operators, that are referred to as "search operators" (or genetic operators), such as recombination and mutation. Each individual in the population receives a measure of its fitness in the environment. Reproduction focuses attention on high fitness individuals, thus exploiting the available fitness information. Recombination and mutation perturb those individuals, providing general heuristics for exploration. Although simplistic, these algorithms are sufficiently complex to provide robust and powerful adaptive search mechanisms.

A **genetic algorithm** (GA) [21] is a model of machine learning inspired by the mechanisms of genetics, which has been applied to optimisation. It operates with an initial population containing a number of trial solutions. Each member of the population is evaluated (to yield a fitness) and a new generation is created from the better of them. The process is continued through a number of generations with the aim that the population should evolve to contain an acceptable solution. In [36] it is stated that GAs are particularly suitable for solving complex optimization problems and hence for applications that require adaptive problem-solving strategies. In order to make genetic algorithms reach an optimal solution faster, parallel implementations of GAs are often used [9].

Genetic algorithms are used for a number of different application areas. An example of this would be multidimensional optimisation problems in which the character string of the chromosome can be used to encode the values for the different parameters being optimised.

In practice, therefore, we can implement this genetic model of computation by having arrays of bits or characters to represent the chromosomes. Simple bit manipulation operations allow the implementation of crossover, mutation as well as other operations. Crossover involves combining strings to swap values, e.g. $101001 + 111111 \rightarrow 101111$. Mutation involves spontaneous alteration of characters in a string, e.g. $000101 \rightarrow 100101$. Although a substantial amount of research has been performed on variable-length strings and other structures, the majority of work with genetic algorithms is focused on fixed-length character strings.

**Statistical classification** is a type of supervised learning algorithm which takes a feature representation of an object or concept and maps it to a classification label. A classification algorithm is designed to learn, or in other words, to approximate the behaviour of a function which maps a vector of features $[X_1, X_2, ..., X_n]$ into one of several classes by looking at several input-output examples of the function.

An instance of a classification algorithm is called a classifier. Learning Classifier Systems [25] are a machine learning technique which combines evolutionary computing and reinforcement learning to produce adaptive systems. It is a minimal form of modelling learning in the sense that it is not necessary to make assumptions about the way the agents perform their reasoning. In addition to that, the absence of any assumptions or biases in the learning process leads to results that can be generalised. A classifier consists of a set of rules, which have a condition C (`if` part) an action A (`then` part) and a strength measure $s$. An example of a classifier system is shown in table II.

| if | $C_1$ | then | $A_1$, | $s_1$ |
|---|---|---|---|---|
| if | $C_2$ | then | $A_2$, | $s_2$ |
| if | $C_3$ | then | $A_3$, | $s_3$ |
| if | ... | then | ... | ... |
| ⋮ | ⋮ | | ⋮ | |

TABLE II

AN EXAMPLE OF A CLASSIFIER SYSTEM.

In the model described in detail in section IV-B, genetic algorithms and classification algorithms have been used to model evolution of populations of solutions and learning.

## REFERENCES

[1] Yaneer Bar-Yam. A mathematical theory of strong emergence using multiscale variety. *Complexity*, 9(6):15–24, 2004.
[2] James Bessen. Open source software: Free provision of complex public goods, 2002. Unpublished working paper, Research on Innovation.
[3] Z. Bodie, A. Kane, and A. Marcus. *Investments, 5th Edition*, chapter 12: Market Efficiency. McGraw-Hill and Irwin, 2002.
[4] Eric Bonabeau, Marco Dorigo, and Guy Theraulaz. *Swarm intelligence: from natural to artificial systems*. Oxford University Press, Inc., New York, NY, USA, 1999.
[5] Andrea Bonaccorsi and Cristina Rossi. Why Open Source software can succeed. *Research Policy*, 32(7):1243–1258, 2003.
[6] Andrea Bonaccorsi and Cristina Rossi. Altruistic individuals, selfish firms? The structure of motivation in Open Source software. *First Monday*, 9(1), 2004.
[7] P.B. (ed.) Brazdil. *Editorial, Machine Learning: Proceedings of the European Conference on Machine Learning*. Springer, New York, NY, USA, 1993.
[8] T. Brenner. *Local Industrial Clusters, Existence, Emergence and Evolution*. Studies in Global Competition. Routledge, London, 2004.
[9] E. Cantu-Paz. A survey of parallel genetic algorithms. *Calculateurs Paralleles, Reseaux et Systems Repartis*, 10(2):141–147, 1998.
[10] K.M. Carley, D.B. Fridsma, E. Casman, A. Yahja, N. Altman, L.-C. Chen, B. Kaminsky, and D. Nave. Biowar: Scalable agent-based model of bioattacks. *IEEE Transactions on Systems, Man and Cybernetics, Part A*, 36(2):252–265, 2006.
[11] M. Chli, P. De Wilde, J. Goossenaerts, V. Abramov, N. Szirbik, L. Correia, P. Mariano, and R. Ribeiro. Stability of multi-agent systems. In *Proceedings of the 2003 IEEE International Conference on Systems, Man, and Cybernetics*, pages 551–556, 2003.
[12] A. Damodaran. *Investment Valuation, 2nd Edition*, chapter 6: Market Efficiency - Theory and Models. Wiley, 2001.
[13] DBE. Annex I - Description of Work, Digital Business Ecosystem. Technical report, 2002.
[14] P. De Wilde. Fuzzy utility and equilibria. *IEEE Transactions on Systems, Man and Cybernetics, Part B*, 34(4):1774–1785, 2004.
[15] P. De Wilde, M. Chli, L. Correia, R. Ribeiro, P. Mariano, V. Abramov, and J. Goossenaerts. Adapting populations of agents. *Lecture Notes in Artificial Intelligence*, 2636:110–124, 2003.
[16] T. Eguchi, K. Hirasawa, J. Hu, and N. Ota. Aa study of evolutionary multiagent models based on symbiosis. *IEEE Transactions on Systems, Man and Cybernetics, Part B*, 36(1):179–193, 2006.

[17] Joshua M. Epstein. Modeling civil violence: An agent-based computational approach. *Proceedings of the National Academy of Sciences, U.S.A.*, 99(10, Supplement 3):7243–7250, 2002.

[18] R. Falcone and C. Castelfranchi. The human in the loop of a delegated agent: the theory of adjustable social autonomy. *IEEE Transactions on Systems, Man and Cybernetics, Part A*, 31(5):406–418, 2001.

[19] E. F. Fama. Efficient capital markets: A review of theory and empirical work. *Journal of finance*, 25:383–417, 1970.

[20] Richard P. Gabriel and Ron Goldman. Open source: Beyond the fairytales, 2002. [Online; accessed 25-May-2005].

[21] David E. Goldberg. *Genetic Algorithms in Search, Optimization and Machine Learning*. Addison-Wesley Longman Publishing Co., Inc., Boston, MA, USA, 1989.

[22] Simon Grand, Georg von Krogh, Dorothy Leonard, and Walter Swap. Resource allocation beyond firm boundaries: A multi-level model for open source innovation. *Long Range Planning*, 37(6):591–610, 2004.

[23] Nathan Griffiths and Michael Luck. Coalition formation through motivation and trust. In *Proceedings of the 2nd international joint conference on Autonomous agents and multiagent systems*, pages 17–24, New York, NY, USA, 2003. ACM Press.

[24] John Hagedoorn. Inter-firm R&D partnerships: an overview of major trends and patterns since 1960. *Research Policy*, 31(4):477–492, 2002.

[25] J. H. Holland. Adaptation. *Progress in Theoretical Biology*, 4:263–293, 1976.

[26] Chun-Che Huang. Using intelligent agents to manage fuzzy business processes. *IEEE Transactions on Systems, Man and Cybernetics, Part A*, 31(6):508–523, 2001.

[27] Z. Jing, E. Billard, and S. Lakshmivarahan. Learning in multilevel games with incomplete information. ii. *IEEE Transactions on Systems, Man and Cybernetics, Part B*, 29(3):340–349, 1999.

[28] Justin Pappas Johnson. Open source software: Private provision of a public good. *Journal of Economics and Management Strategy*, 11(4):637–662, 2002.

[29] J. O. Kephart. Software agents and the route to information economy. *Proceedings of the National Academy of Sciences, U.S.A.*, 99(10, Supplement 3):7207–7213, 2002.

[30] Jeffrey O. Kephart, James E. Hanson, and Jakka Sairamesh. Price and niche wars in a free-market economy of software agents. *Artificial. Life*, 4(1):1–23, 1997.

[31] Alan P. Kirman and Nicolaas J. Vriend. Evolving market structure: An ACE model of price dispersion and loyalty. *Journal of Economic Dynamics and Control*, 25(3):459–502, 2001.

[32] Blake LeBaron. Building the Santa Fe Artificial Stock Market. Working Paper, June 2002. Available at http://www.econ.iastate.edu/tesfatsi/blake.sfisum.pdf.

[33] Josh Lerner and Jean Tirole. Some simple economics of open source. *Journal of Industrial Economics*, 50:197–234, June 2002.

[34] Christoph H. Loch and Bernardo A. Huberman. A punctuated-equilibrium model of technology diffusion. *Management Science*, 45(2):160–177, 1999.

[35] M. Luck, P. McBurney, and C. Preist. *Agent Technology: Enabling Next Generation Computing (A Roadmap for Agent Based Computing)*. AgentLink, 2003.

[36] J. L. Ribeiro Filho, P. C. Treleaven, and C. Alippi. Genetic-algorithm programming environments. *Computer*, 27(6):28–43, 1994.

[37] R. Savit, R. Manuca, and R. Riolo. Adaptive competition, market efficiency and phase transitions. *Physical Review Letters*, 82(10):2203–2206, 1999.

[38] Klaus Schmidt and Monika Schnitzer. Public Subsidies for Open Source? Some Economic Policy Issues of the Software Market. Technical Report 3793, C.E.P.R. Discussion Papers, February 2003. Available at http://ideas.repec.org/p/cpr/ceprdp/3793.html.

[39] Kwang Mong Sim and Eric Wong. Toward market-driven agents for electronic auction. *IEEE Transactions on Systems, Man and Cybernetics, Part A*, 31(6):474–484, 2001.

[40] M. Sysi-Aho, A. Chakraborti, and K. Kaski. Searching for good strategies in adaptive minority games. *Physical Review E*, 69(3):36125–1–36125–7, 2004.

[41] Leigh Tesfatsion. *Handbook of Computational Economics, Vol. 2: Agent-Based Computational Economics*, chapter 1, Agent-Based Computational Economics: A Constructive Approach to Economic Theory. North-Holland, 2005. To appear.

[42] H. Van Dyke Parunak, Sven Brueckner, and Robert Savit. Universality in multi-agent systems. In *AAMAS '04: Proceedings of the 3rd International Joint Conference on Autonomous Agents and Multiagent Systems*, pages 930–937, Washington, DC, USA, 2004. IEEE Computer Society.

[43] T. Yamasaki and T. Ushio. An application of a computational ecology model to a routing method in computer networks. *IEEE Transactions on Systems, Man and Cybernetics, Part B*, 32(1):99–106, 2002.